\documentstyle[11pt,epsf]{article}
\begin{document}
%%%%%%%%%%%%%%%%%%%%%%%%%%%%%%%%%%%%%%%%%
\baselineskip=\normalbaselineskip\multiply\baselineskip 
by 150\divide\baselineskip by 100
\pagenumbering{arabic}
\pagestyle{plain}
\reversemarginpar
%%%%%%%%%%%%%%%%%%%%%%%%%%%%%%%%%%%%%%%%%
\newcommand{\gsim}{\lower.7ex\hbox{$\;\stackrel{\textstyle>}{\sim}\;$}}
\newcommand{\lsim}{\lower.7ex\hbox{$\;\stackrel{\textstyle<}{\sim}\;$}}
\renewcommand{\thefootnote}{\fnsymbol{footnote}}
\setcounter{footnote}{1}
%%%%%%%%%%%%%%%%%%%%%%%%%%%%%%%%%%%%%%%%%
\begin{titlepage}
\begin{flushright}
ANL--HEP--PR--96--43\\
UM--TH--96--06\\
\end{flushright}
\begin{flushright}
May 1996
\end{flushright}
\vspace{0.4cm}
\begin{center}
\large
{\bf Do About Half the Top Quarks at FNAL Come From\\
Gluino Decays?}
\end{center}
\begin{center}
{\bf G.~L. Kane}\footnote{gkane@umich.edu}
\end{center}
\begin{center}
{Randall Laboratory of Physics \\
University of Michigan\\
Ann Arbor, MI 48104}
\end{center}
\begin{center}
and
\end{center}
\begin{center}
{\bf S. Mrenna}\footnote{mrenna@hep.anl.gov}
\end{center}
\begin{center}
{High Energy Physics Division\\
Argonne National Laboratory \\
Argonne, IL  60439}
\end{center}
\vspace{0.4cm}
\raggedbottom
\setcounter{page}{1}
\relax

\begin{abstract}
\noindent
We argue that it is possible to make a consistent picture of FNAL data
including the production and decay of gluinos and squarks.
The additional cross section is several pb, about the size of that for Standard
Model (SM) top quark
pair production.  If the stop squark mass is small enough, about
half of the top quarks decay to stop squarks, and the loss of SM
top quark pair production rate is compensated by the supersymmetric
processes.  This behavior is consistent with the reported top quark
decay rates in various modes
and other aspects of the data,
and suggests several other possible decay signatures.
This picture can be tested easily with more data, perhaps even with
the data in hand, and 
demonstrates the potential power of a hadron collider to
determine supersymmetric parameters.   It also has implications for
the top mass measurement and the interpretation of the LEP $R_b$ excess.
\\ \\ PAC codes: 12.60.Jv, 12.15.Mm, 14.65.Ha, 14.80.L
\end{abstract}

\end{titlepage}
\newpage
%%%%%%%%%%%%%%%%%%%%%%%%%%%%%%%%%%%%%%%%%
\renewcommand{\thefootnote}{\arabic{footnote}}
\setcounter{footnote}{0}
%%%%%%%%%%%%%%%%%%%%%%%%%%%%%%%%%%%%%%%%%

\section{Introduction}
\indent

While there is still no compelling experimental evidence that nature is
supersymmetric on the weak scale, there have been recent reports of data
that encourage this view.  The most explicit is an event in
CDF~\cite{park} that does not have a probable SM interpretation, and can
naturally be explained as selectron pair production\cite{sandro,dine}.
In the interpretation when the lightest neutralino is the LSP,
the analysis of this event leads to a fairly well determined range of
masses and couplings for sleptons $\tilde \ell$, charginos $C_i$, and 
neutralinos $N_i$.  
$C_i$ and $N_i$ are the chargino and
neutralino mass eigenstates, with $C_1,N_1$ ($C_2,N_4$) being the 
lightest (heaviest) chargino and neutralino.
If in
addition there is a light stop squark $\tilde t$
(i.e., $m_{\tilde t} \lsim m_W$),
it is remarkable that the chargino mass and couplings from~\cite{sandro}
can explain~\cite{wells} the LEP reported excess for $Z\to b\bar b$
decays $(R_b)$, at least if that excess is not too large.  Another
encouraging result~\cite{kane} is that the LSP resulting from these
studies has a mass and coupling such that it is a good candidate for the
cold dark matter of the universe, giving $0.1 \lsim \Omega_{\rm LSP} h^2 
\lsim 1.$  In the following, we use the $N_i$ and $C_i$ masses and couplings
reported in Ref.~\cite{sandro}.

If the stop squark $\tilde t$ and at least one neutralino $N_i$ are light, 
the top quark decay $t\to \tilde t N_i$ must occur along with 
$t\to bW$ with branching ratio about one--half.  
Currently, the FNAL top quark pair production counting rate is interpreted as
a measurement of $\sigma(p{\bar p}\rightarrow t{\bar t}X) 
\times {\rm BR}^2 (t\to bW).$ 
The reported values from CDF for this are about 7 pb \cite{CDF}, 
which is already larger than the
predicted total cross section~\cite{berger} (about 5.5 pb) for the top quark
mass $(m_t)$ extracted from kinematic reconstruction (about 175 GeV).  
The D0 measurement for the production rate is smaller, but 
suffers from larger backgrounds.
Unless there is additional production of top quarks or
other particles with similar decay modes,
there is no room for extra decay modes, such as $t\to \tilde
t N_i$, in this data sample (this is made quantitative below).

In this paper we observe that supersymmetry,
with certain reasonable and well--motivated choices of sparticle masses,
will lead to extra top quark production.
Additionally, there are other final states from sparticles
with a high purity of
$b$--quarks, leptons, and jets; they can mimic the top quark signal.
If squark ($\tilde q$) and gluino ($\tilde g$) masses 
(excepting the stop squark) are between
200 and 300 GeV, then they have pb--level production rates at the Tevatron.
We assume squarks are heavier than gluinos, motivated by the
results of Ref.~\cite{sandro} as well as the observation that
gluinos decay predominantly to top quarks and stop squarks if
the decays to other squarks are not kinematically allowed.
A lower bound is set for the gluino mass by requiring
$m_{\tilde g} > m_t + m_{\tilde t}$ so that the decay
${\tilde g} \to t{\tilde t}^{*}$ can occur for physical masses.
Since gluinos are Majorana particles, they will decay
equally to $t {\tilde t}^{*}$ and $\bar t \tilde t$ \cite{barnett}.
Pair production of $\tilde g \tilde g$ will give a final state containing
$tt, \bar t\bar t,$ and $t\bar t$ in the ratio $1:1:2$.
If squarks are heavier than gluinos, they decay as $\tilde q \to
\tilde g q,$ $C_i q, N_i q$; $\tilde g q$ has the largest coupling but
the smallest phase space.  
We assume that $R$--parity is
approximately conserved, as implied by the interpretation of the
CDF event as sparticle
production, so that all sparticles decay to $N_1$ within the detector,
and $N_1$ subsequently
escapes.  The couplings to $C_i$ and $N_i$ are
largely determined by the analysis of Ref.~\cite{sandro}.  The size of
the BR$(\tilde q \to \tilde g q)$ is determined by the available
phase space.  We freely use the term Minimal Supersymmetric Standard
Model (MSSM) to describe our models, and it should be understood that
we do not assume gaugino mass unification, but we do assume that all
squarks except the stop squark are mass degenerate.

The relative masses of $\tilde t$ and $C_i$ are important for the
signatures, and are not determined yet.  Motivated by the LEP $R_b$
data, we
assume $m_{\tilde t} \lsim m_{C_i}.$ Then the dominant decays are $C_i
\to b{\tilde t}^{*},$ and $\tilde t \to c N_i$~\cite{rudaz2}.
If half of
the top quarks decay to stop squarks, 
then $\bar t(\to{\bar b}W) t(\to \tilde t N_i)$ followed
by $\tilde t\to cN_j,$ so half of all $t \bar t$ events give a $Wbc$
signature; the finite detector acceptance, clustering of jets, 
etc., might lead to a similar final state in the Standard Model,
but at a much lower rate.
Similar observable differences
occur in every distinct $t\bar t$ channel; 
results for some examples are shown in
Sec.~4.  Furthermore, a different value of $m_t$ is likely to be
extracted from 
different channels if it is determined by comparison with a SM Monte
Carlo, and the true value of $m_t$ might not be the apparent one.
Note that a large number of charm jets arise from stop decays; if
they could be tagged, e.g. by the lepton in charm semi--leptonic decays,
it would help test our arguments.

In the following section, we review several
results on the top quark which are relevant to our discussion.
In Section~3, we present a supersymmetric
model which is consistent with the results of Section~2, 
explains several other
pieces of data, and predicts several signatures in the present and
future FNAL data samples.  The detailed study of these models
are shown in Section~4.  Finally, Section~5 contains our conclusions.

\section{Upper Bound on
Non--SM Decay Modes of The Top Quark} 
\indent

Based on a number of measurements and predictions, it is possible
to bound non--SM decay modes of the top quark which
are ``invisible'' to the standard searches.  The final state
resulting from $t{\bar t}\to bW {\bar c}N_1N_1$ or $t{\bar t}\to
cN_1N_1{\bar c}N_1N_1$ would not have enough leptons or jets to be
included in the leptonic, dileptonic, or hadronic event samples, and,
hence, is invisible. 
Here, we review briefly a previous analysis bounding these ``invisible''
decays\cite{yuan}.

The FNAL experiments report essentially two
independent measures of the top quark mass:  the kinematic measure,
whereby the four--vectors of all the decay products are reconstructed
into the four--vector of the parent particle, and the counting measure,
whereby one compares the observed number of events, corrected by
efficiencies, to the production cross section as a function of mass.
Additionally, based on single-- and
double--$b$--tagged events, CDF has reported a measurement of
BR$(t\to qW)$ for $q \not= b.$~\cite{incandela}.  We interpret this
as a limit on $b_W \equiv$ BR$(t \to bW)$.
The SM prediction for 
the production cross section $\sigma_t$ is bounded theoretically 
and the current best prediction of $\sigma_t$
at $m_t = 175$ GeV is 5.52$^{+.07}_{-.45}$ pb from the first reference 
of~\cite{berger}.
Given the measured quantities and the theoretical predictions 
and their uncertainties, one can
perform a $\chi^{2}$ minimization in the variables $m_t$, $\sigma_t$,
and $b_W$.
Finding the minimum value $\chi^2_{min}$ yields $m_t =
168.6^{+3.0}_{-3.0}$ GeV, $\sigma_{t\overline{t}} =
7.09^{+.68}_{-.62}$\,pb and $b_W = 1.00^{+.00}_{-.13}$.
At the 95\% confidence level, $b_W \geq .74$, so an upper limit 
on BR($t\rightarrow X$), where $X \ne bW$, exists of about 25\%.

This analysis disfavors a large component of non--SM top
quark decays, such as a $\tilde t\to c N_1.$  However, this result
has limitations.  Namely, it does not include the possibility that
the same physics which allows new top quark decays can also lead to
more top quark production.  
(See
\cite{sender} for an analysis which includes $\tilde t 
{\tilde t}^{*}$ production
and the decays $\tilde t \to b C_1$ and concludes that a light 
$\tilde t$ is
not excluded by the present FNAL data even for BR($t\to{\tilde t}N_i$)=1/2).
Furthermore, it assumes that all of the
new decay modes result in final states which elude the
standard searches.  Finally, 
it does not include the
LEP indirect fits to $m_t$ which favor a lower value (for
example, the world average top mass including LEP, DIS, SLC, and
FNAL data is 161$\pm$8 GeV assuming an 80 GeV Higgs boson, as
expected from Ref.~\cite{wells}).
Since
this would weight the $\chi^2$ for a 
smaller $m_t$ and a larger $\sigma_t$, $b_W$ would be smaller.
and more ``invisible'' decays would be allowed. 
Based on these observations, it is premature to conclude that
a light stop is inconsistent with the observed top quark events.
In the following section, we present explicit supersymmetric models
motivated by Ref.~\cite{sandro}
which successfully address several of the loopholes in the previous analysis.
Squarks and gluinos are produced at a significant rate and decay 
largely into top quarks.  The stop squark is light
enough that the charginos $C_1$ and $C_2$ decay to $b{\tilde t}^{*}$,
which in combination with other decays, give top--like final states.
Finally, a lower top mass predicts a larger SM production rate,
allowing more ``invisible'' decays.

\section{Top Quarks from Squark and Gluino Decay}
\indent

If selectrons, charginos, and neutralinos have masses of order $m_Z$,
then squarks and gluinos might be light enough to be produced
in significant numbers at FNAL.  The analysis of Ref.~\cite{sandro} is
done with a general low energy softly broken supersymmetry theory,
without assumptions about gaugino or squark mass unification.  
As a result, the gluino and squark masses are not determined.  
However, there are phenomenological reasons to settle on the range
200--300 GeV
for these masses.
Given the analysis of $R_b$, we expect a light stop squark
with mass less than about $M_W$.  LEP and FNAL limits do not allow this to
be too small \cite{claes}, and we are forced to $m_{\tilde t}$ values 
somewhere between 
about $M_Z/2$ and $M_W$.  As explained in the previous section,
light stop squarks alone dilute the signal $t\to bW$ through
decays $t\to \tilde t N_i$ at a level incompatible with the data, 
so a new top quark or top--like production mechanism is needed. 
The simplest method is to use the decay channel 
${\tilde g}\to t{\tilde t}^{*}$, which requires
$m_{\tilde g} > m_t + m_{\tilde t}$.
For this to be the dominant decay channel, the other
squarks must be heavier than the gluino.
Finally, the Tevatron has limited parton
luminosities to produce heavy particles.  Since about half of the
top quark decays ``disappear'', we need a production mechanism which
is about the same size as the SM rate.  Therefore, we
are naturally lead to squark and gluinos masses between 200 and 300 GeV.
We will see that a number of observables depend on the particular masses,
so eventually they can be directly measured.

A similar mass hierarchy follows from theoretical considerations.
Ref.~\cite{sandro} found gaugino masses obeying the mass relations
$M_1 \simeq M_2,$ rather than
the unification relation $M_1 \simeq {1\over 2} M_2.$  This could be
explained by anomalous behavior of the U(1) mass, so that the non-Abelian
masses may still approximately satisfy the 
unification relation $M_2 \simeq M_3.$
Then the gluino mass should be about three times the $C_1$ mass, or in
the range 195--270 GeV.  Similarly, we assume here that squarks are
(except for stop mass eigenstates that are expected to be separated)
approximately degenerate, and about 2.5 times the selectron mass, as
suggested by models.  For numerical work, we take a common
squark mass $m_{\tilde q}$ 
for left- and right-handed squarks of five families which is
slightly above $m_{\tilde g}$.
We study models with $160<m_t<175$ GeV, $210<m_{\tilde g}<235$ GeV,
$220<m_{\tilde q}<250$ GeV, and $45<m_{\tilde t}<60$ GeV.
All results are based on the
analyses and models of ref.~\cite{sandro,wells} and are thus consistent
with existing evidence for supersymmetry  and with other particle
physics constraints.

\section{Numerical Results}
\indent

In this section, we present separate results on the counting measurement
of the top production cross section and
kinematic measurement of the top quark mass in the supersymmetric models
described in the previous section.  All event simulation is performed
using the Monte Carlo {\tt PYTHIA 5.7} with supersymmetric extensions
\cite{spythia}.  The cross sections are computed at the Born level with
additional QCD radiation added in the leading--log approximation.
The structure functions used are CTEQ2L, and the hard--scattering
scale used in the evaluation of the structure functions and the
running couplings is the partonic center of mass energy 
(transverse mass) if the partonic final state has no (some) 
non--zero QCD quantum numbers.  Particle energies are smeared
using Gaussian resolutions based on the CDF detector, and jets are defined
using the {\tt PYTHIA LUCELL} subroutine.  Jets are
$b$--tagged with a constant efficiency when they contain a 
high--$p_T$ $b$--parton; the exact efficiency is stated only
when a result depends upon the choice.  
All leptons and photons must
be isolated from excess transverse energy.

To eliminate dependence on the particulars of $b$--tagging and isolation
efficiencies and detector cracks, we will present some results
as ratios with the SM signal expected using the same
Monte Carlo routines.  
Where there is no SM signal expected, or just a small one, then
we show an expected number of events in 100 pb$^{-1}$.  
First, we examine the counting measurement of the
top quark production cross section.  
Since there are so many production processes and
decay chains at work when all sparticles have weak scale masses, 
it is clear that a multi--channel
analysis should be performed.  
To illustrate this point, we present a Unitarity Table (Table 1), based on
a very loose set of experimental cuts, so that almost 100\% of the
simulated events fall into some category.  
On one hand, this is useful to check
that a particular model does not contradict data by predicting an
anomalously high rate in some channel.  On the other
hand, it can point the way to unexpected features in the data, or 
explain why various signals are not present.  This particular table
is for a representative model, but displays the essential features.
The columns represent
various production mechanisms:  (1) SM $t\bar t$,
(2) $t\bar t$ with supersymmetric decays, 
(3) squark pair, (4) squark--gluino,
(5) gluino pair, and (6) $C_i$ or $N_i$ in association with a squark or gluino.
The rows represent various final states defined by the presence of
charged leptons $\ell^\pm$, photons $\gamma$, and a large missing
transverse energy (${\hbox to -2pt{/\hss}}E_T$) 
and the number of jets $n_j$ (if present, $\ell^\pm$
and $\gamma$ are explicitly noted) :  
(a) $n_j < 3$,
(b) $n_j\ge 3$, with small ${\hbox to -2pt{/\hss}}E_T$, 
(c) $n_j\ge 3$, with large ${\hbox to -2pt{/\hss}}E_T$, 
(d) the subset of (b) and (c) which has $n_j\ge 6$, 
(e) $\ell^\pm, n_j<3$, a lepton with less than 3 jets, 
(f) $\ell^\pm, n_j\ge 3$ and large ${\hbox to -2pt{/\hss}}E_T$,
(g) $\ell^\pm, n_j\ge 3$ and small ${\hbox to -2pt{/\hss}}E_T$, 
(h) $\ell^\pm\ell^\pm,n_j\ge 0$ with large ${\hbox to -2pt{/\hss}}E_T$, 
(i) $\ell^\pm\gamma, n_j\ge 0$,
(j) $\gamma,n_j\ge 0$, and 
(k) $\gamma\gamma, n_j\ge 0$.
The numbers in each column represent the fraction of generated events
in a particular final state.  If the most important final states are
accounted for, then the sum for each column should be close to unity,
excluding the 6 jet final state, which is a subset of two others.  
Indeed, the sums vary between .98 and 1.00.
This table demonstrates that we understand where the individual
supersymmetric contributions go.  Only after multiplying by
the production cross section (and choosing more realistic cuts)
for each process can we determine the
observable rate.  
Row (c) corresponds to the standard SUSY search mode of
multi--jets plus ${\hbox to -2pt{/\hss}}E_T$.  
For the particular choices of squark and gluino
masses considered here, we still elude the present experimental bound.

\begin{table}
\renewcommand{\arraystretch}{1.33}
\begin{center}
\begin{tabular}[bht]{|c||c|c|c|c|c|c|}\hline
Mode & SM $t\overline{t}$ & MSSM $t\overline{t}$ &
$\tilde{q}\tilde{q}$ & $\tilde{q}\tilde{g}$ &
$\tilde{g}\tilde{g}$ & $\tilde{q}\tilde{\chi}$ \\ \hline\hline 
$n_j<3$           & -- & .17 & .02 & .06 & .12 & .17 \\ \hline
$n_j\ge 3, {\hbox to -2pt{/\hss}}E_T <$ 60 GeV 
                  & .59 & .33 & .20 & .27 & .36 & .31 \\ \hline
$n_j\ge 3, {\hbox to -2pt{/\hss}}E_T >$ 60 GeV 
                  & .10 & .21 & .25 & .25 & .25 & .15 \\ \hline
($n_j\ge 6$)      &(.42)&(.14)&(.12)&(.18)&(.21)&(.06)\\ \hline
$\ell^\pm, n_j<3$ & .01 & .08 & .01 & .02 & .05 & .05 \\ \hline
$\ell^\pm, n_j\ge 3, {\hbox to -2pt{/\hss}}E_T >$ 25 GeV 
                  & .21 & .11 & .08 & .10 & .12 & .06 \\ \hline
$\ell^\pm, n_j\ge 3, {\hbox to -2pt{/\hss}}E_T <$ 25 GeV 
                  & .04 & .02 & .01 & .01 & .02 & .02 \\ \hline
$\ell\ell, {\hbox to -2pt{/\hss}}E_T > $25 GeV 
                  & .03 & .01 & .01 & .01 & .01 & .01 \\ \hline
$\ell^\pm\gamma$  & --  & .01 &.04 & .03 & .01  & .03 \\ \hline
$\gamma$          & --  & .06 &.31 & .22 & .05  & .18 \\ \hline
$\gamma\gamma$    & --  & --  &.06 & .01  & --  & .02  \\
\hline
\end{tabular}
\end{center}
\caption{Unitarity Table, illustrating the fraction of events which
fall into several categories for SM $t\bar t$ and MSSM processes.}
\end{table}

In the following, we present results based on more realistic
cuts.  There are three standard top quark search modes defined by
the CDF cuts: (i) leptonic, (ii) dileptonic, and (iii) hadronic.
We find substantial signals in 2 additional channels:
(iv) $"W"bc$, and (v) $\gamma bc$.  The channel (iv) cuts are the same
as for (i), except only 2 jets are allowed.  
The excess number of $"W"bc$ events (where $"W"$ may have a
different transverse mass since $N_1$'s carry away energy) would
appear as an excess of $W$ plus 2 jet events with one $b$--tag; the
second jet is charm which can be tagged with a lower efficiency.
The channel (v)
cuts require a high--$p_t$ $\gamma$ ($>20$ GeV) in the central rapidity
region ($|\eta^\gamma|<1$), with 2 or more additional jets.  One of the
two leading jets must have a $b$-tag.  As discussed earlier,
channel (iv) should be limited in the Standard Model.  Channel (v) events
arise from the decay $N_2\to N_1\gamma$ in association with top quark
or top--like decays, and should have a tiny contribution from 
$t \bar t$ production alone in the Standard Model.

Table~2 summarizes the results of the counting experiments
for the various models considered.  The numbers have ranges, and we
comment on the correlations later.
In the upper portion of the table, we present
fractions in each row of
our Monte Carlo estimate of the process rate after cuts 
divided by the same estimate
for SM top quark production.
These numbers suggest that different values will be obtained for
the top production cross section in different modes; in a given mode,
the value will depend on the analysis and cuts.  This is consistent
with the reported CDF and D0 cross sections\cite{CDF}.
The row labelled $\ell^\pm\ell^\pm$ shows the predicted non--SM
signal of like sign leptons; this is possible because of the Majorana
nature of the gluino.  We note that about 1/7--1/5 of all
dilepton events should have
leptons with the same charge.
The middle section of the table shows the expected number of events
in 100 pb$^{-1}$ for the two aforementioned channels $"W"bc$ and
$\gamma bj$.  These numbers do not include a $b$--tagging efficiency.
Also included in the $\gamma bj$ sample is the expected number of 
events from $C_i(\to b\tilde t )N_2(\to N_1\gamma)$ production (+35).
The final section shows the variation in total production
cross sections for the various channels.  Note that the MSSM $t\bar t$
production cross section is identical to the SM one; the only difference
occurs in the allowed top quark decays.
Of course, the various apparent cross sections are correlated.
For $m_t$=160 GeV, $m_{\tilde{q}}$=220 GeV, $m_{\tilde{g}}$=210 GeV, 
$m_{\chi^0_1}$=38 GeV, $m_{\tilde{t}_1}$=45 GeV, the cross sections
measured in the three modes are 6.5, 7.8, and 6.6 pb.  
For $m_t$=165 GeV, $m_{\tilde{q}}$=240 GeV, $m_{\tilde{g}}$=220 GeV, 
$m_{\chi^0_1}$=38 GeV, $m_{\tilde{t}_1}$=50 GeV, the numbers 
are 5.7, 5.3, and 6.3 pb.

\begin{table}
\renewcommand{\arraystretch}{1.33}
\begin{center}
\begin{tabular}[bht]{|c||c|c|c|c|c||c||c|}\hline
Mode & MSSM $t\overline{t}$ &
$\tilde{q}\tilde{q}$ & $\tilde{q}\tilde{g}$ &
$\tilde{g}\tilde{g}$ & $\tilde{q}\tilde{\chi}$ & MSSM sum &
``$\sigma_{t\overline{t}}$''\\ \hline\hline 
\multicolumn{8}{|l|}{Ratio with expected SM cross section}\\
\hline\hline
$\ell^\pm n_j\ge 3$& .35--.43 & .07--.10 & .13--.19 & 
.05--.06 & .03--.04 & .71--.74 & 3.9--6.5
\\ \hline	   
$\ell^\pm\ell^\mp$ &.31--.38 & .10--.21 & .08--0.18 & .03--.05 & 
.04--.05 & .58-.87 & 3.8--7.8
\\ \hline	   
$\ell^\pm\ell^\pm$ &.03--.04 & .02--.04 & .02--.03 & .02--.03 & 
.01--.01 & .10--.14 & \\ \hline
$n_j\ge 6$         &.28--.35 & .10--.16 & .20--.23 & .07--.08 & 
.01--.03 & .66--.86 & 4.3--6.5
\\ \hline\hline	   
\multicolumn{8}{|l|}{Number of expected events in 100 pb$^{-1}$}\\
\hline\hline
$\ell^\pm n_j=2$   &13--19 & 0--2 & 1--5 & 1--3 & 2--4 & 17--33 & \\
$\gamma bj$        &7--13 &6--22 &6--23 & 0--2 & 5--9 & 26--69+35 & \\
\hline\hline	   
\multicolumn{8}{|l|}{Production Cross Sections}\\
\hline\hline
Total $\sigma$ (pb)& 5.5--9.0 & 1.7--4.1 & 1.9--5.2 & 0.6--1.5 & 
1.0--1.8 & 10.7--21.6  & \\
per channel        &      &      &      &      &      &   & \\
\hline
\end{tabular}
\end{center}
\caption{Expected results of the top quark counting experiments
for the MSSM.  The apparent top production cross sections are
shown in the final column.  The number of events in the present
data sample for two channels are displayed in the middle section.
Typical MSSM production cross sections appear in the final section.}
\end{table}

Some of the larger apparent rate for top quark production
for SUSY processes comes from the increased cross 
section for smaller $m_t$.  We use the resummed prediction,
which is similar to the NLO number.  The production of squarks and
gluinos, on the other hand, are calculated only at LO, and could 
receive a substantial NLO correction.  
Based on Ref.~\cite{zerwas}, we estimate that our squark and gluino
production cross sections (which are evaluated using the transverse
mass as the factorization scale) could be increased (but not decreased)
by as much as 40\%.  We have not included this $K$--factor in our
study.
In addition, smaller $m_t$
allows for smaller gluino and squark masses, which further increases
the MSSM rate.  We have made no attempt to optimize the numbers
in Table~2.  It is remarkable how naturally the
apparent cross section values span the experimentally allowed values.

Since they are a novel feature of our models,
we also show a typical scatter plot of the events 
expected in 100 pb$^{-1}$ with
signature $\gamma b$ ${\hbox to -2pt{/\hss}}E_T$+jets in Fig.~1, 
resulting from models with $m_t$=160 GeV.
They are of particular interest because
there is no parton--level SM source of such events, and our MSSM
scenario predicts a significant number.  There are three sources:
$q \bar q \to C_i(\to b{\tilde t}^{*})N_2(\to\gamma N_1),
q \bar q(gg) \to t(\to bW(\to jj)){\bar t}(\to {\tilde t}^{*}N_2(\to\gamma
N_1)),$ 
with
${\tilde t}\to cN_1$,
and cascade decays from $\tilde q\tilde q$, $\tilde q\tilde g$,
$\tilde g\tilde g$, $\tilde g N_i, \tilde g C_i$, $\tilde q N_i, \tilde
q C_i$,  populating different regions of the plot.
The $C_i N_2$ and $t\bar t$ signals depend mostly on the supersymmetric
interpretation of the CDF event and the postulate of a light stop squark
to explain $R_b$; their signal may be present regardless of the 
other squark and gluino masses.
Note that the first of these produces only two prompt jets, while the
other two produce several jets.
Finding these events could confirm supersymmetry in general and
our arguments in particular.
%%%%%%%%%%%%%%%%%%%%%%%%%%%%%%%%%%%%%%%%%%%%%%%%%%%%%%%%%%%%%%%%%%%%%
\begin{figure}
\centering
\epsfxsize=3.5in
\hspace*{0in}
\epsffile{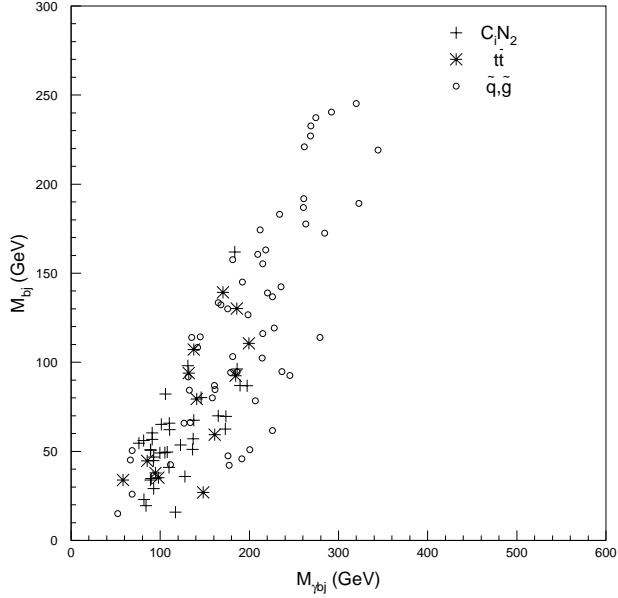}
\vspace*{.5in}
\caption{The distribution of $\gamma bj$ events expected in 100 
pb$^{-1}$.  There are contributions from the $C_i N_2$,
$t\bar t$, and $\tilde q,\tilde g$ production processes.}
\end{figure}
%%%%%%%%%%%%%%%%%%%%%%%%%%%%%%%%%%%%%%%%%%%%%%%%%%%%%%%%%%%%%%%%%%%%%

%%%%%%%%%%%%%%%%%%%%%%%%%%%%%%%%%%%%%%%%%%%%%%%%%%%%%%%%%%%%%%%%%%%%%
\begin{figure}
\begin{center}
\epsfxsize=3.5in
\hspace*{0in}
\epsffile{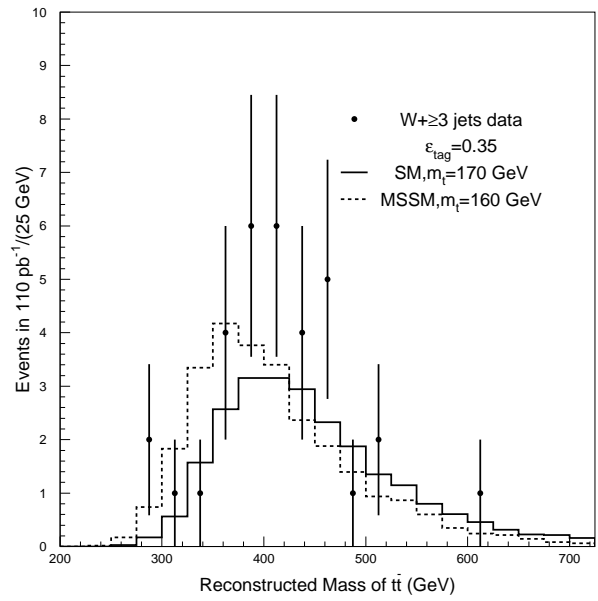}
\vspace*{.5in}
\end{center}
\caption{The invariant mass distribution of the $t\bar t$ pair
for SM and MSSM top quark production compared with the CDF data.}
\end{figure}
%%%%%%%%%%%%%%%%%%%%%%%%%%%%%%%%%%%%%%%%%%%%%%%%%%%%%%%%%%%%%%%%%%%%%

In addition to the top quark measurements based on counting events,
there are kinematic measurements, such as 
the reconstructed top mass, the $t\bar t$ invariant mass, the 
transverse momentum of the pair, etc.
The explicit reconstruction is a difficult task.  Not all of the apparent
top quark signal in our models comes from real top quarks, but a
substantial fraction does.  Rather than attempt an explicit reconstruction
for such detector dependent quantities,
we have tried to determine where we should expect 
agreement with and deviation from the SM distributions.  We identify the
partonic top and anti--top quarks in our generated events 
and use these
to calculate the invariant mass and transverse
momentum of the $t\bar t$ pair.  In the process, we ignore those events
which have only one or no real top quarks, but scale those events with
two top quarks to the full rate including the discarded events.
The invariant mass distribution from SM top quark production alone with
$m_t$=170 GeV, for our MSSM with $m_t$=160 GeV, and with the
CDF data is displayed in Fig.~2. The CDF data is from \cite{tartarelli}.
While we have only performed a crude simulation, we believe this figure
demonstrates that our MSSM is consistent with the data.
This consistency is understandable.
Since top 
quarks are coming from gluino decay just above threshold, they are
produced almost at rest in the lab frame.  As a result, the distribution
must peak slightly above $2m_t$.  

The transverse momentum of the $t \bar t$ pair is displayed 
in Fig.~3 for the same conditions as for Fig.~2.  Here, we have not
smeared the distributions, but we observe that the MSSM distribution
is broader than that expected in the SM.
%%%%%%%%%%%%%%%%%%%%%%%%%%%%%%%%%%%%%%%%%%%%%%%%%%%%%%%%%%%%%%%%%%%%%
\begin{figure}
\begin{center}
\epsfxsize=3.5in
\hspace*{0in}
\epsffile{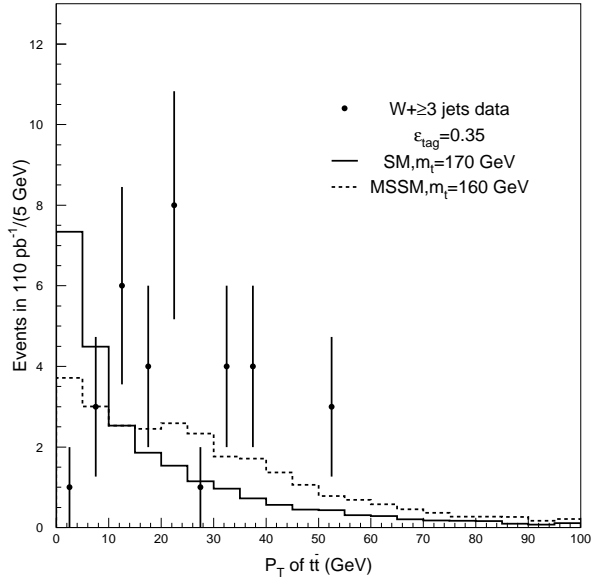}
\vspace*{-.5in}
\end{center}
\caption{The transverse momentum distribution of the $t\bar t$ pair
for the SM and MSSM compared to the data.  The generated distributions
have not been smeared.}
\end{figure}
%%%%%%%%%%%%%%%%%%%%%%%%%%%%%%%%%%%%%%%%%%%%%%%%%%%%%%%%%%%%%%%%%%%%%
This
too is expected; the two--body decay of the gluinos
tends to randomize the top quark direction, removing the approximate
balance in $p_T$ expected from SM $t \bar t$ production.
The expected SM
$t\bar t$ distribution displayed in Ref.~\cite{tartarelli} is
narrower than the data, so the MSSM could explain this discrepancy.

The events used for the kinematic reconstruction of $m_t$ with a
$b$--tag come from  the $W(\to \ell\nu)$+jets mode.  
The lepton and the ${\hbox to -2pt{/\hss}}E_T$ in these
events should have a transverse mass consistent with that from the
decay of a $W$ boson.  When top quarks are produced in MSSM events,
there can be additional ${\hbox to -2pt{/\hss}}E_T$ from cascade decays down 
to $N_1$.
The expected distributions for the transverse mass and the CDF data
are shown in Fig.~4.  
The naive expectation that the MSSM distribution must be
distinguishably different is not fulfilled.
The CDF data is from \cite{transmass}.
%%%%%%%%%%%%%%%%%%%%%%%%%%%%%%%%%%%%%%%%%%%%%%%%%%%%%%%%%%%%%%%%%%%%%
\begin{figure}
\begin{center}
\epsfxsize=3.5in
\hspace*{0in}
\epsffile{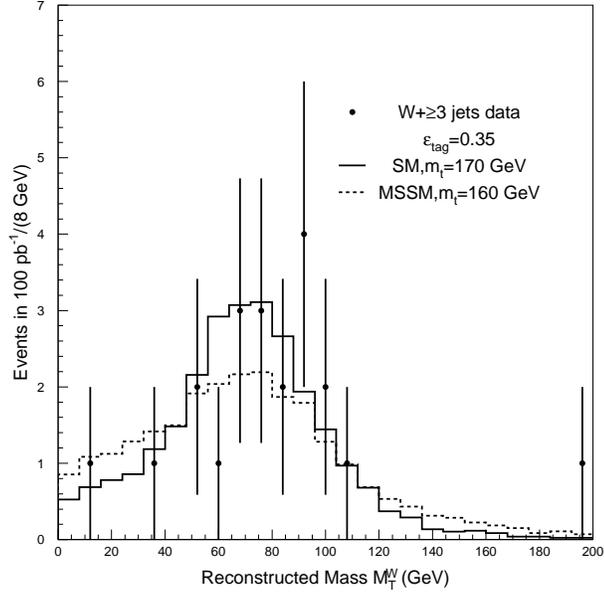}
\vspace*{.5in}
\end{center}
\caption{The transverse mass of the lepton and ${\hbox to -2pt{/\hss}}E_T$ for
the SM and MSSM compared to the data.}
\end{figure}
%%%%%%%%%%%%%%%%%%%%%%%%%%%%%%%%%%%%%%%%%%%%%%%%%%%%%%%%%%%%%%%%%%%%%

Since many of the top events have associated jets, the apparent top
mass deduced from such events will only be the actual top mass if
very particular cuts and analyses are used.  For example, we have noticed 
that the invariant mass distribution of the leptons in dilepton events
is softer than for SM top events.  This indicates that the mass
kinematically reconstructed from dilepton events will be lower in the
MSSM than 
for the other modes.  Note that this is the only mode which does
not require a $b$--tag, so that the additional jets from squark
decays can enhance the signal.

\section{Conclusions}
\indent

We have argued that existing data is consistent with the possibility
that hundreds of squarks and gluinos have been produced at FNAL.
Squarks decay mainly into gluinos, charginos, and neutralinos, gluinos
into top quarks and stop squarks, 
and BR($t\to \tilde t N_i$) is about 1/2.  We have
checked that the predicted counting measures and kinematic measures
are consistent with the available data, and, in some cases, give a
better description.  A number
of associated predictions allow this view to be tested, possibly with
existing data.  If correct, it has implications for the top quark mass
and cross section measurements, for interpreting the LEP $R_b$ data,
and of course for the existence of supersymmetry in nature.

\section*{ Acknowledgments }
\indent

The authors thank E.L.~Berger, H.~Frisch,  E.~Kovacs,
T.~LeCompte, L.~Nodulman, M.~Strovink,
S.~Ambrosanio, G.D.~Kribs and S.P.~Martin 
for useful discussions.

\end{document}